\documentclass{emulateapj}
\usepackage{times}
\usepackage{apjfonts}
\usepackage{amssymb}
\usepackage{epsfig}

\usepackage{graphicx,graphics}
\usepackage{epsfig}
\usepackage{dcolumn}
\usepackage{bm}
\usepackage{natbib}
\usepackage{times}
\usepackage{ulem}
\def\kmm#1  {{\bf [KMM:~ #1]~}}
\def\new#1 {{\bf #1 }}
\def\cut#1 {\sout{#1} }

\newcommand{\hi}{H~{\sc i}}
\newcommand{\pmn}{PMN~J0134$-$0931}
\newcommand{\pks}{PKS~1413+135}

\newcommand{\dal}{\ensuremath{\lsb \Delta \alpha/ \alpha \rsb}}

\newcommand{\dmu}{\ensuremath{\lsb \Delta \mu/\mu \rsb}}
\newcommand{\beq}{\begin{equation}}
\newcommand{\eeq}{\end{equation}}

\newcommand{\lsb}{\left[}
\newcommand{\rsb}{\right]}

\shorttitle{Fundamental constant evolution with conjugate-satellite OH lines}
\shortauthors{Kanekar et al. }
\begin{document}

\title{Probing fundamental constant evolution with redshifted conjugate-satellite OH lines}

\author{Nissim Kanekar\altaffilmark{1}, Jayaram N. Chengalur\altaffilmark{2}, Tapasi Ghosh\altaffilmark{3}}
\altaffiltext{1}{Ramanujan Fellow, National Centre for Radio Astrophysics, TIFR, Pune 411 007, India; 
National Radio Astronomy Observatory, 1003 Lopezville Road, Socorro, NM 87801, USA; nkanekar@ncra.tifr.res.in}
\altaffiltext{2}{National Centre for Radio Astrophysics, TIFR, Pune 411 007, India}
\altaffiltext{3}{Arecibo Observatory, HC03 Box 53995, Arecibo, PR 00612, USA}

\begin{abstract}

We report Westerbork Synthesis Radio Telescope and Arecibo Telescope observations 
of the redshifted satellite OH~18cm lines at $z \sim 0.247$ towards PKS~1413+135. The 
``conjugate'' nature of these lines, with one line in emission and the other in 
absorption, but with the same shape, implies that the lines arise in the same gas. 
The satellite OH~18cm line frequencies also have different dependences on the 
fine structure constant $\alpha$, the proton-electron mass ratio $\mu = m_p/m_e$, 
and the proton gyromagnetic ratio $g_p$.  Comparisons between the satellite line redshifts 
in conjugate systems can hence be used to probe changes in $\alpha$, $\mu$, and $g_p$,
with few systematic effects.  The technique yields the expected null result when 
applied to Cen.A, a nearby conjugate satellite system.  For the $z \sim 0.247$ 
system towards PKS~1413+135, we find, on combining results from the two telescopes, 
that $\left[ \Delta G/G \right] = (-1.18 \pm 0.46) \times 10^{-5}$ (weighted mean), 
where $G = g_p \left[ \mu \alpha^2 \right]^{1.85}$; this is tentative evidence 
(with $2.6 \sigma$ significance, or at 99.1\% confidence) for a smaller value 
of $\alpha$, $\mu$, and/or $g_p$ at $z \sim 0.247$, i.e. at a lookback time of 
$\sim 2.9$~Gyrs. If we assume that the dominant change is in $\alpha$, this 
implies $\left[ \Delta \alpha / \alpha \right] = (-3.1 \pm 1.2) \times 10^{-6}$. 
We find no evidence that the observed offset might be produced by systematic 
effects, either due to observational or analysis procedures, or local conditions 
in the molecular cloud.

\end{abstract}

\section{Introduction} 
\label{sec:intro}

Most modern higher-dimensional theories venturing beyond
the standard model of particle physics contain the fairly generic feature that 
the low-energy fundamental constants should vary with time. The detection 
of such changes opens up an avenue to probe basic physics, especially 
important because other predictions of these models tend to lie at unattainably 
high energies ($\gtrsim 10^{19}$~GeV).  Tests of such low-energy predictions 
of these models, such as changes in the constants, violation of the 
equivalence principles, etc, may provide the only means of distinguishing 
between different unified theories and are hence of much importance in physics.

Astrophysical techniques can be used to probe changes in fundamental
constants like the fine structure constant $\alpha$, the proton-electron
mass ratio $\mu \equiv m_p/m_e$ and the proton gyromagnetic ratio $g_p$, 
over a large fraction of the age of the Universe (e.g. 
\citealp{savedoff56,wolfe76,thompson75,dzuba99,chengalur03}). Indeed, 
\citet{murphy04} applied one such technique (the ``many-multiplet method''; 
\citealp{dzuba99}) to High Resolution Echelle Spectrograph (HIRES) data from 
the Keck telescope to obtain $\dal = (-5.7 \pm 1.1) \times 10^{-6}$ from 143~absorbers 
at $0.2 < z < 4.2$, i.e. suggesting a smaller value of $\alpha$ at earlier times
(see also \citealp{murphy03}). This result has not so far been confirmed with 
independent data on smaller samples from the Ultraviolet Echelle Spectrograph 
(UVES) on the Very Large Telescope (e.g. \citealp{molaro08,srianand07b,murphy08b}.
More recently, \citet{king08} used redshifted molecular hydrogen (H$_2$) lines 
to place strong constraints on changes in $\mu$, obtaining $\dmu = (-2.6 \pm 3.0) 
\times 10^{-6}$ from three absorbers at $z \sim 2.8$. Note that these results are all based on optical spectra, 
where wavelength calibration, line blending, intrinsic velocity offsets, isotopic 
abundances, interloping absorbers, etc, are all possible sources of systematic error, 
(e.g. \citealp{murphy03,griest10}). Further, the quoted errors in the result of 
\citet{murphy04} do not include effects from systematic distortions in the 
HIRES wavelength calibration \citep{griest10}; similar, but smaller, distortions have 
also been found in the wavelength scale of the VLT-UVES spectrograph \citep{centurion09,whitmore10}.

Given the possibility that underestimated or unknown systematics might dominate the 
errors from a given technique, it is important that independent techniques, with 
entirely different systematics, be used to test for changes in the constants. 
It is also crucial to probe evolution at all timescales, as the timescales of 
the putative changes are entirely unknown. Redshifted radio OH lines provide 
an independent approach to study changes in $\alpha$, $\mu$ and $g_p$ 
\citep{chengalur03,darling03,kanekar04a}. In rare cases, the satellite OH~18cm
lines (at rest frequencies of 1612.230825 (15)~MHz and 1720.529887 (10)~MHz; 
\citealp{lev06}) are ``conjugate'' to each other; i.e. the lines have the same shapes, but 
one line is in absorption and the other in emission. This masing effect 
arises due to the quantum mechanical selection rules for decay routes to 
the {$2\Pi_{3/2}$ (J = 3/2)}~OH ground state, after the molecules have been 
pumped to higher excited rotational states \citep{elitzur92}. Crucially, 
the conjugate behavior {\it ensures that the satellite lines arise from 
the same gas}. These lines are thus well-suited for measuring changes in $\alpha$, 
$\mu$ and $g_p$ between the source redshift and today, as systematic velocity 
offsets between the lines are ruled out by the maser mechanism. Any observed 
difference between the line redshifts must then arise due to a change in 
one or more of the above constants \citep{kanekar08b}.

Only two conjugate OH systems are known at cosmological distances, at $z \sim 
0.247$ towards \pks\ \citep{kanekar04b,darling04} and at $z \sim 0.765$ towards 
\pmn\ \citep{kanekar05}. We report here on deep Westerbork Synthesis Radio Telescope 
(WSRT) and Arecibo Telescope (AO) OH observations of \pks, that yield tentative
evidence for changes in $\alpha$, $\mu$ and/or $g_p$ over a lookback
time of $\sim 2.9$~Gyr.

\begin{figure*}
\epsfig{file=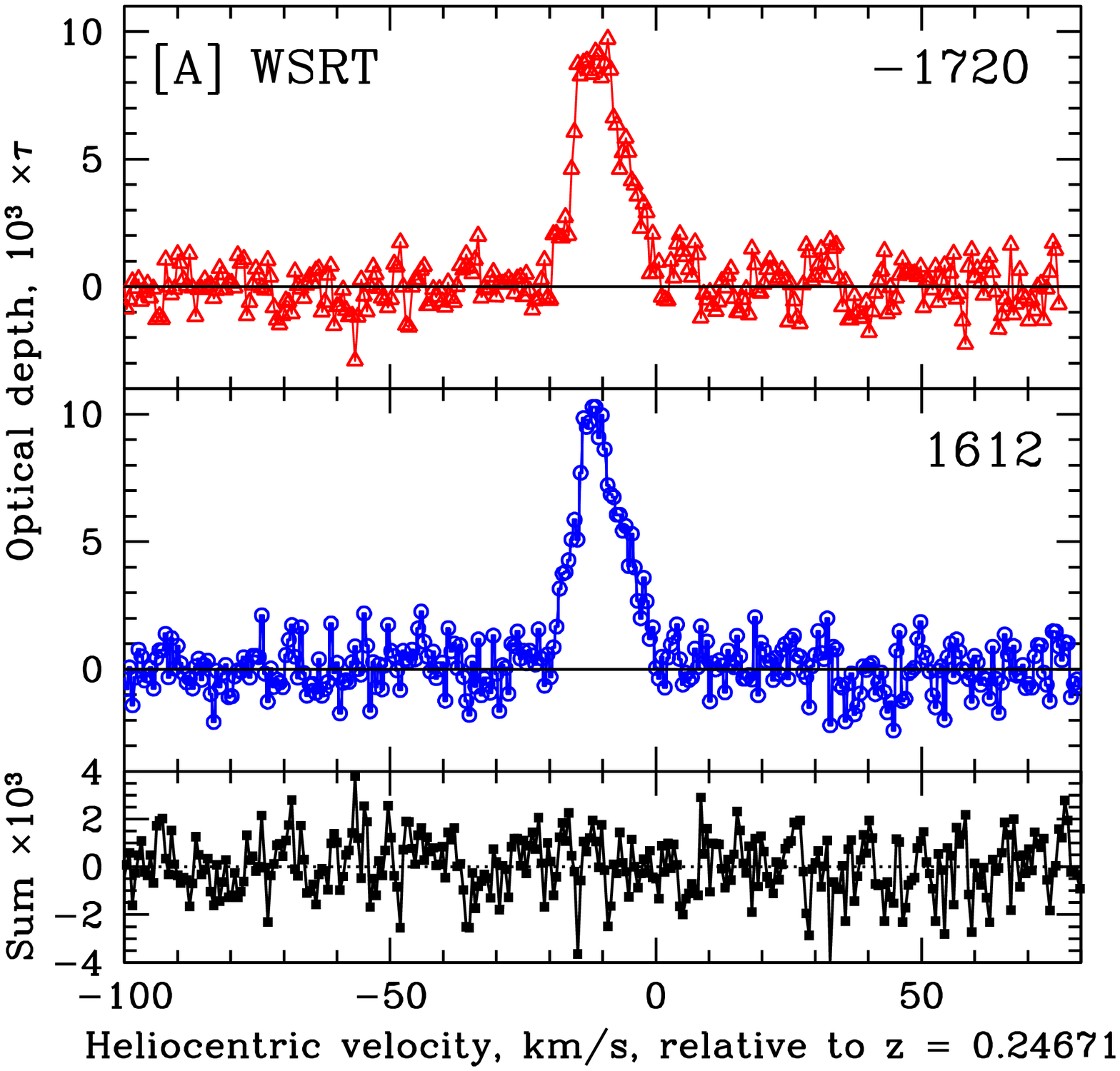,width=3.5in,height=3.5in}
\epsfig{file=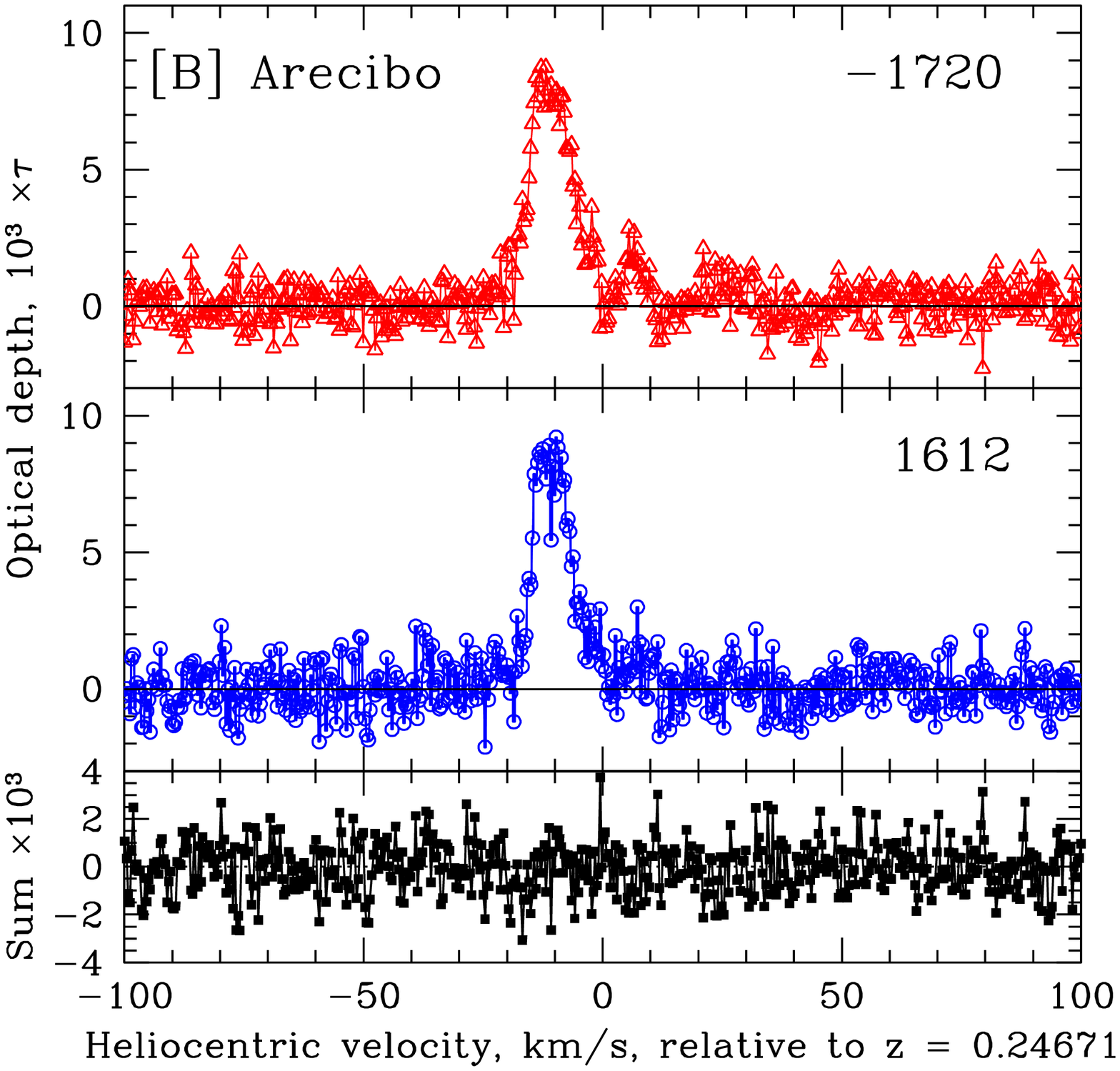,width=3.5in,height=3.5in}
\caption{Top panels: WSRT and AO satellite OH spectra towards \pks, with optical 
	depth ($1000 \times \tau$) plotted against heliocentric velocity, 
	in km/s, relative to $z = 0.24671$. The 1612 and 1720~MHz spectra 
	are plotted using open circles and triangles, respectively, with the 1720~MHz 
	spectrum flipped in sign. Bottom panels: The sum of the 1612 and 1720~MHz optical 
	depth spectra from each telescope. }
\label{fig:fig1}
\end{figure*}

\section{Spectra and results} 
\label{sec:spectra}

The WSRT and the AO were used to carry out deep integrations on the satellite OH 
lines of \pks\ between May and July~2005 (WSRT; $\sim 58$~hours) and June~2008
and May~2009 (AO; $\sim 40$~hours in double-position-switched mode; \citep{ghosh02}). 
The 1720~MHz and 1612~MHz lines 
were observed simultaneously in all runs, with velocity resolutions of $\sim 0.35$~km/s (AO) 
and $\sim 0.57$~km/s (WSRT), after Hanning smoothing. The WSRT data were analysed in 
``classic'' {\sc AIPS}, while the Arecibo data were analysed in IDL, both using standard 
procedures.

The top four panels of Fig.~\ref{fig:fig1}[A] and [B] show the satellite OH~18cm optical depth 
spectra from each telescope, with the 1720~MHz spectrum flipped in sign for the comparison.
The WSRT spectra of Fig.~\ref{fig:fig1}[A] have root-mean-square (RMS) optical depth noise 
values of $8.7 \times 10^{-4}$~(1612)  and $8.3 \times 10^{-4}$~(1720), per $\sim 0.57$~km/s 
channel, while the corresponding RMS noise values for the AO spectra of Fig.~\ref{fig:fig1}[B] 
are $4.6 \times 10^{-4}$~(1612) and $4.4 \times 10^{-4}$~(1720), per $\sim 0.35$~km/s channel. 
The bottom panels of the figure show the sum of the 1612 and 1720~MHz optical depth profiles 
from each telescope. The RMS noise values on the summed spectra
are $\sim 1.2 \times 10^{-3}$ per $\sim 0.57$~km/s channel (WSRT) and $\sim 6.3 \times 10^{-4}$ 
per $\sim 0.35$~km/s channel (AO), consistent with the noise levels on the individual 
spectra. The summed spectra also show no evidence for non-Gaussian structure; indeed, 
a Kolmogorov-Smirnov (K-S) rank-1 test finds that the summed spectra are consistent (at
$\le 1.3\sigma$ significance) with being drawn from normal distributions. The satellite 
OH lines are thus conjugate at the signal-to-noise ratio of our observations. 

Having confirmed that the satellite lines are conjugate within our measurement errors, the 
next step is to test whether they arise at the same redshift. For lines that 
are not conjugate, this is usually done by fitting Gaussians (or Voigt profiles) to each 
line, to measure the redshift of peak absorption (e.g. \citealp{murphy03,srianand07b,kanekar05}).
In the case of the WSRT and AO spectra, a three-component Gaussian model provides a good
fit to each spectrum, yielding noise-like residuals. However, the process of fitting 
multiple spectral components to a spectral line of shape can itself affect the results, 
especially in the case of complex profiles. An important advantage of using conjugate 
satellite OH lines to probe fundamental constant evolution is that the shapes of the 
lines are known to be the same. One can thus directly
determine the peak of the cross-correlation of the two lines, instead of having to decompose
each profile into its components. The cross-correlation of the two WSRT satellite OH 
spectra was found to peak at $\Delta V = (-0.37 \pm 0.22)$~km/s, and that of the AO spectra 
at $\Delta V = (-0.20 \pm 0.10)$~km/s. The RMS noise values were estimated by 
cross-correlating $10^4$ pairs of simulated spectra. Each simulated spectrum was
obtained by adding independent representations of Gaussian random noise to the 
best 3-component fit, with the noise spectra characterized by the RMS values of the observed 
spectra. A weighted average of the WSRT and AO results gives a net velocity offset of 
$(-0.23 \pm 0.09)$~km/s between the two satellite lines, with the 1720~MHz line at a 
higher velocity. 

Note that the sum of the optical depth spectra might be expected to approach zero 
more closely (i.e. to be more ``noise-like'') on applying the measured offset between 
the line profiles. At present, the fact that the measured offset is smaller than a 
pixel ($\sim$ half a pixel for the Arecibo spectra) implies that this test is not 
very sensitive, especially because the offset is detected at relatively low 
statistical significance. Note, further, that a K-S test finds that the summed optical 
depth spectra, obtained without applying the velocity offset, are consistent with 
noise (at $\le 1.3\sigma$ significance). However, we did carry out the above test on 
the higher-sensitivity Arecibo spectra, and found, as expected, that no 
statistically-significant difference is apparent in either the mean or the RMS 
noise of the summed spectrum, on applying the measured velocity offset before
carrying out the sum. Finally, we emphasize that this would be a useful test to 
carry out on higher-sensitivity, higher-resolution data, to test for an offset 
of multiple pixels. This test would also verify that the two lines indeed have 
precisely the same shape, thus complementing the cross-correlation technique,
which formally only measures the offset between the two profiles, assuming that
they have the same shape.

The above agreement between the shapes (amplitudes and widths) of the satellite 
lines of Fig.~\ref{fig:fig1} and the offset between the line redshifts is precisely 
the signature of an evolution in the fundamental constants that was being sought. 
Using the equations of \citet{chengalur03}, the velocity offsets measured in the cross-correlation 
analysis yield $\lsb \Delta G/G \rsb = (-1.9 \pm 1.1) \times 10^{-5}$ for the WSRT 
spectra, and $\lsb \Delta G/G \rsb = (-1.03 \pm 0.51) \times 10^{-5}$ for the AO spectra, 
where $G \equiv g_p \left[ \mu \alpha^2 \right]^{1.85}$. A weighted average of the WSRT 
and AO results gives $\lsb \Delta G/G \rsb = (-1.18 \pm 0.46) \times 10^{-5}$. This 
is weak ($\sim 2.6 \sigma$) evidence for a change in $\alpha$, $\mu$ and/or $g_p$, 
from the redshift of \pks\ to the present epoch, i.e. over 2.9~Gyrs.

We also carried out a similar cross-correlation analysis on the strong conjugate satellite 
OH lines from the nearby extra-galactic source Cen.A \citep{langevelde95}, to test whether 
velocity offsets between the satellite lines are seen in local conjugate systems. The analysed 
OH spectra towards this source were kindly provided to us by Huib van Langevelde. The peak of 
the cross-correlation for the Cen.A spectra was found to be at 
$\Delta V = (+0.05 \pm 0.11)$~km/s, consistent with no velocity offset between the lines. 
This demonstrates that the conjugate satellite technique provides the expected null result 
for at least one local system, and at a sensitivity comparable to that of the datasets towards 
\pks.

\section{Systematic effects}
\label{sec:systematics}

Systematic effects that might contribute to increased errors, 
above those determined from the cross-correlation analysis, stem from two possible sources. 
``Local'' issues that could result in an observed offset include errors in the telescope 
frequency scale, doppler-tracking issues, terrestrial radio frequency interference (RFI), 
etc. However, both the WSRT and AO data were 
recorded in topocentric frequency, with no corrections made for Earth motion during 
the observations.  These corrections were applied during the data analysis, using a 
model of Earth motion accurate to $< 15$~m/s, an order of magnitude lower than our 
measurement errors. Frequency 
calibration is also not usually an issue in radio spectroscopy, with the frequency scale 
set by the accuracy of masers and local oscillators (typically a few~Hz, two orders of 
magnitude lower than our errors). The laboratory OH frequencies are known to an 
accuracy of $\sim 15$~Hz \citep{lev06}, again far smaller than our errors. Finally, there is
no evidence for non-Gaussian structure in the baselines of the different spectra, that
might limit their dynamic range. The only ``local'' source of systematic error in this 
technique thus appears to be RFI; the WSRT and AO data were carefully edited
and there was no evidence of residual RFI in any of the spectra. Further, note that 
a consistent velocity offset between the satellite lines is found in both the WSRT 
and AO spectra, although the two datasets were acquired with telescopes on different 
continents and at different times; this makes it very unlikely that the offset might be 
caused by RFI.

Another category of systematics consists of ``astronomical'' effects that might cause 
intrinsic velocity offsets between the two lines (e.g. different ``clouds'' excited in 
each line, source inhomogeneities, an interloping line from a different transition, etc). 
However, unlike in the optical regime, there is no possibility of ``confusing'' lines 
from \pks\ or an interloper along the line of sight. All possible line interlopers at 
these frequencies are weak and only detectable in very high {\sc HI} column density gas, 
requiring a sightline through a galaxy. The relatively-large AO beam intersects three known 
galaxies, the Milky Way, the host galaxy of \pks\ (at $z \sim 0.24671$) and 
SDSS~J141557.20+131851.2, a galaxy at $z \sim 0.3687$; there are no transitions from any 
of these redshifts that might be blended with the OH satellite lines. For example, the 
nearest known Galactic lines are those of CH$_2$CHCN, at a frequency of $\sim 1371.7$~MHz, 
$\sim 8.3$~MHz away from the redshifted 1720~MHz frequency, while the closest known
lines from $z \sim 0.24671$ are those of CH$_3$OCHO, $\sim 1.0$~MHz away from the redshifted 
1612~MHz line frequency. Finally, we note that the separation between transitions 
from different OH isotopes at these frequencies is $\gtrsim 50$~MHz, much larger than 
our observing bandwidth. 

The strongest argument against intrinsic velocity offsets, RFI, or additional spectral 
lines, playing a significant role is the fact that the shapes of the satellite lines are 
in excellent agreement, in both WSRT and AO datasets. This is a crucial difference between 
the present technique, based on the conjugate satellite OH lines, and other approaches. 
Here, the observed lines themselves provide a stringent test of their use as a probe 
of fundamental constant evolution: the maser mechanism giving rise to the lines \citep{elitzur92} 
implies that {\it the line shapes must agree} if they arise in the same gas. 

It should also be pointed out that the satellite line shapes only agree at our 
present sensitivity. An additional weak spectral component might be present in one of 
the two spectra, below our detection threshold; this could result in the cross-correlation 
peak being offset from zero velocity, while retaining the conjugate behaviour. While 
we cannot as yet formally rule out this possibility, it should be emphasized that 
this can easily be tested by deeper spectra in the satellite OH lines.


\section{Discussion}
\label{sec:which}

If the observed ($\sim 2.6 \sigma$ 
significance) offset between the satellite lines is not caused by a ``hidden'' spectral 
component (or RFI) below our sensitivity threshold, it could be evidence for changes 
in one or  more of $\alpha$, $\mu$, or $g_p$. Note that the conjugate satellite technique
is directly sensitive to changes in $G \equiv g_p \left[ \mu \alpha^2 \right]^{1.85}$, 
and cannot provide additional information on which constant is changing. Our 
present result $\Delta G/G = (-1.18 \pm 0.46) \times 10^{-5}$ implies the relation 
$3.7\dal + 1.85\dmu + [\Delta g_p/g_p] = (-1.18 \pm 0.46) \times 10^{-5}$, suggesting
that one or more of $\alpha$, $\mu$ and $g_p$ had smaller values at $z = 0.247$ than
at the present epoch. The limiting cases, assuming that only one of $\alpha$, $\mu$, and 
$g_p$ changes, are $\dal = (-3.1 \pm 1.2) \times 10^{-6}$ (for $\dal >> \dmu$, $[\Delta g_p/g_p]$), 
$\dmu = (-6.2 \pm 2.4) \times 10^{-6}$  (for $\dmu >> \dal$, $[\Delta g_p/g_p]$), and
$[\Delta g_p/g_p] = (-1.18 \pm 0.46) \times 10^{-5}$ (for $[\Delta g_p/g_p] >> \dal$, $\dmu$).

Changes in $g_p$ are expected to be far smaller than those in $\mu$ and $\alpha$ 
\citep{langacker02}.  Further, a stringent constraint on changes in $\mu$ from a 
higher redshift absorber has recently been obtained from comparisons between NH$_3$ inversion and 
HCN/HCO$^+$ rotational lines: \citet{murphy08} find $\dmu < 1.6 \times 10^{-6}$ ($2\sigma$) 
from a gravitational lens at $z \sim 0.685$. This suggests that changes in 
$G \equiv g_p [\mu \alpha^2]^{1.85}$ at $z \sim 0.25$ may be dominated by changes 
in $\alpha$, i.e. the first
limiting case above, yielding $\dal = (-3.1 \pm 1.2) \times 10^{-6}$. It is interesting 
that this result for $\dal$ is similar in both amplitude and sign to the value of $\dal$ 
obtained by \citet{murphy04} using the many-multiplet method at a higher redshift, 
$\dal = (-5.7 \pm 1.1) \times 10^{-6}$ at $\langle z \rangle = 1.75$. 

Other sensitive astronomical techniques to probe fundamental constant evolution 
include those based on the many-multiplet method \citep{murphy04}, H$_2$ lines \citep{king08}, 
\hi~21cm and OH lines \citep{kanekar05}, inversion and rotational lines 
\citep{flambaum07b}, \hi~21cm and ultraviolet carbon lines \citep{kanekar10}, etc. 
All of these are affected by systematic effects of different 
types (e.g. \citealp{kanekar08b}). For example, \citet{griest10} have recently shown
that Keck-HIRES spectra are affected by wavelength calibration errors of unknown origin;
the effect of these systematic errors on the many-multiplet result of \citet{murphy04} 
is not known, although it seems clear that the error budget will increase. Similar 
problems have been found with VLT-UVES spectra (e.g. \citealp{centurion09,whitmore10}), which 
would affect the H$_2$ results of \citet{king08}. 

The importance of the conjugate satellite OH technique stems from the fact that 
it has fewer systematic effects than other approaches and allows a test of its 
own applicability, in the prediction that the shapes of the satellite OH lines 
must be the same. We have also demonstrated that the expected null result is 
obtained on applying the technique to a local conjugate satellite system, Cen.A. 
Equally important, and again unlike most other techniques, the conjugate 
satellite lines allow an estimate of changes in the fundamental constants from 
a {\it single space-time location}, without any averaging over multiple absorbers 
(which, for example, is essential in the many-multiplet method to average out 
local systematics). The only apparent drawback to the technique is the possibility of
RFI at the line frequencies; this can be ameliorated by using multiple 
telescopes, as done here, or multiple observing epochs, taking advantage of the 
shift in the line frequency in the terrestrial frame due to the motion of the Earth 
around the Sun.

In summary, we have obtained deep WSRT and Arecibo Telescope spectra in the redshifted 
satellite OH~18cm lines from \pks, at $z \sim 0.247$. The lines are 
conjugate at our sensitivity, with the sum of the optical depths consistent 
with noise. We find that the peak of the cross-correlation of the satellite OH~18cm 
lines is offset from zero velocity (at $\sim 2.6 \sigma$ significance), tentative 
evidence for a redshift offset between them. We have 
examined the data for systematic effects and find no evidence that such effects
might give rise to the observed velocity offset. We have also 
verified that the cross-correlation of the satellite OH~18cm lines in the local 
conjugate satellite system in Cen.A peaks at zero velocity, within the errors.
The observed offset in satellite line redshifts in \pks\ implies $\Delta G/G = 
(-1.18 \pm 0.46) \times 10^{-5}$, where $G \equiv g_p [\mu \alpha^2]^{1.85}$, 
suggesting that one or more of $\alpha$, $\mu$, and $g_p$ had smaller values 
at $z \sim 0.247$ than at the present epoch.


\acknowledgments
We thank Huib van Langevelde for providing us the conjugate satellite spectra 
towards Cen.A, Carlos Martins for stimulating discussions, and Rene Vermeulen and Chris 
Salter for much help with the WSRT and AO observations. The WSRT is operated 
by the ASTRON (Netherlands Foundation for Research in Astronomy) with support from 
the Netherlands Foundation for Scientific Research NWO. The Arecibo Observatory is a part 
of the NAIC, operated by Cornell University under a cooperative agreement with the NSF. 
NK acknowledges support from an NRAO Jansky Fellowship and a Ramanujan Fellowship, 
and from the Max Planck Society.

\bibliographystyle{apj}

\end{document}